\documentclass[12pt,a4paper]{article}
\usepackage[dvips]{graphics}

\begin{document}
\renewcommand{\thefootnote}{\fnsymbol{footnote}}
\newcommand{\epem}{\mbox{$\mathrm{e^+e^-}$}}
\newcommand{\dbar}{\mbox{$\overline{\mathrm{d}}$}}
\newcommand{\sbar}{\mbox{$\overline{\mathrm{s}}$}}
\newcommand{\lamsb}{\mbox{$\Lambda_{\overline{MS}}$}}
\newcommand{\lamsbsq}{\mbox{$\Lambda_{\overline{MS}}^2$}}
\newcommand{\amz}{\mbox{$\alpha_s(M_{\mathrm{Z}^0})$}}
\newcommand{\mz}{\mbox{$M_{\mathrm{Z}^0}$}}
\newcommand{\as}{\mbox{$\alpha_{\mathrm{s}}$}}
\newcommand{\assq}{\mbox{$\alpha_{\mathrm{s}}^2$}}
\newcommand{\asb}{\mbox{$\overline{\alpha}_{\mathrm{s}}$}}
\newcommand{\asbsq}{\mbox{$\overline{\alpha}_{\mathrm{s}}^2$}}
\newcommand{\ash}{\mbox{$\widehat{\alpha}_{\mathrm{s}}$}}
\newcommand{\oa}{\mbox{${\cal O}(\alpha_{\mathrm{s}})$}}
\newcommand{\oaa}{\mbox{${\cal O}(\alpha_{\mathrm{s}}^2)$}}
\newcommand{\oaaa}{\mbox{${\cal O}(\alpha_{\mathrm{s}}^3)$}}
\newcommand{\Zzero}{\mbox{${\mathrm{Z}^0}$}}
\newcommand{\ecm}{\mbox{$E_{cm}$}}
\newcommand{\musq}{\mbox{$\mu^2$}}
\newcommand{\yc}{\mbox{$y_{cut}$}}
\newcommand{\xmu}{\mbox{$x_{\mu}$}}
\newcommand{\xmusq}{\mbox{$x_{\mu}^2$}}
\newcommand{\mht}{\mbox{$M_H$}}
\newcommand{\bt}{\mbox{$B_T$}}
\newcommand{\bw}{\mbox{$B_W$}}
\newcommand{\CF}{\mbox{$C_{\mathrm{F}}$}}
\newcommand{\dd}{\mbox{$\mathrm{d}$}}
\newcommand{\chisq}{\mbox{$\chi^2$/d.o.f.}}
\newcommand{\Opal}{\mbox{O{\sc pal}}}
\newcommand{\Aleph}{\mbox{A{\sc leph}}}
\newcommand{\Delphi}{\mbox{D{\sc elphi}}}
\newcommand{\Jetset}{\mbox{J{\sc etset}}}
\newcommand{\Herwig}{\mbox{H{\sc erwig}}}
\newcommand{\Ariadne}{\mbox{A{\sc riadne}}}
\newcommand{\Cojets}{\mbox{C{\sc ojets}}}
\newcommand{\boldn}{\mbox{\boldmath$n$}}
\newcommand{\boldp}{\mbox{\boldmath$p$}}
\newcommand{\costh}{\mbox{$\cos\theta$}}
\newcommand{\qcosth}{\mbox{$q\cdot\cos\theta$}}
\newcommand{\thetaT}{\mbox{$\theta_\mathrm{Th}$}}
\newcommand{\sigmaT}{\mbox{$\sigma_\mathrm{T}$}}
\newcommand{\sigmaL}{\mbox{$\sigma_\mathrm{L}$}}
\newcommand{\sigmatot}{\mbox{$\sigma_\mathrm{tot}$}}
\newcommand{\costhT}{\mbox{$\cos\theta_\mathrm{Th}$}}
\newcommand{\half}{\mbox{$\textstyle\frac{1}{2}$}}
\newcommand{\onethird}{\mbox{$\textstyle\frac{1}{3}$}}
\newcommand{\twothirds}{\mbox{$\textstyle\frac{2}{3}$}}
\newcommand{\threequarters}{\mbox{$\textstyle\frac{3}{4}$}}
\newcommand{\threeeighths}{\mbox{$\textstyle\frac{3}{8}$}}
\begin{titlepage}
%
\begin{center}\mbox{\large\bf EUROPEAN LABORATORY FOR PARTICLE PHYSICS}
\end{center}
\bigskip
\begin{tabbing}
\` CERN--EP/98--124\\
\` July 20, 1998 \\
\end{tabbing}
\begin{center}
\mbox{\LARGE\bf Measurement of the Longitudinal Cross-Section}
\mbox{\LARGE\bf using the Direction of the Thrust Axis}
\mbox{\LARGE\bf in Hadronic Events at LEP}
\end{center}\bigskip
\begin{center}
{\Large \bf OPAL Collaboration}
\end{center}\bigskip   
\bigskip\bigskip\bigskip\bigskip
\begin{center}{\Large\bf Abstract}
\end{center}
In the process $\epem \rightarrow \mathrm{hadrons}$, 
one of the effects of gluon emission is to modify the $(1+\cos^2\theta)$ form
of the angular distribution of the thrust axis, an effect which 
may be quantified by the {\em longitudinal cross-section}.  
Using the \Opal\ detector at LEP, 
we have determined the longitudinal to total cross-section ratio 
to be $\sigmaL/\sigmatot=0.0127\pm0.0016\pm0.0013$ at the parton level, 
in good agreement with the expectation of QCD computed to \oaa.
Comparisions at the hadron level with  
Monte Carlo models are presented.
The dependence of the longitudinal cross-section
on the value of thrust has also been studied, and provides a
new test of QCD.  
\medskip
\noindent
 
\bigskip\bigskip\bigskip
\medskip
\begin{center} 
 
\end{center}

\bigskip\bigskip
\begin{center} 
\end{center}
\bigskip\bigskip
\begin{center} 
 
{\bf Submitted to Phys. Lett. B }
\end{center}
 
\end{titlepage}
\begin{center}{\Large        The OPAL Collaboration
}\end{center}\bigskip
\begin{center}{
G.\thinspace Abbiendi$^{  2}$,
K.\thinspace Ackerstaff$^{  8}$,
G.\thinspace Alexander$^{ 23}$,
J.\thinspace Allison$^{ 16}$,
N.\thinspace Altekamp$^{  5}$,
K.J.\thinspace Anderson$^{  9}$,
S.\thinspace Anderson$^{ 12}$,
S.\thinspace Arcelli$^{ 17}$,
S.\thinspace Asai$^{ 24}$,
S.F.\thinspace Ashby$^{  1}$
% ...
}\end{center}\bigskip
\bigskip
$^{  1}$School of Physics and Astronomy, University of Birmingham,
Birmingham B15 2TT, UK
\newline
$^{  2}$Dipartimento di Fisica dell' Universit\`a di Bologna and INFN,
I-40126 Bologna, Italy
\newline
$^{  3}$Physikalisches Institut, Universit\"at Bonn,
D-53115 Bonn, Germany
\newline
$^{  4}$Department of Physics, University of California,
Riverside CA 92521, USA
\newline
$^{  5}$Cavendish Laboratory, Cambridge CB3 0HE, UK
\newline
$^{  6}$Ottawa-Carleton Institute for Physics,
Department of Physics, Carleton University,
Ottawa, Ontario K1S 5B6, Canada
\newline
$^{  7}$Centre for Research in Particle Physics,
Carleton University, Ottawa, Ontario K1S 5B6, Canada
\newline
$^{  8}$CERN, European Organisation for Particle Physics,
CH-1211 Geneva 23, Switzerland
\newline
$^{  9}$Enrico Fermi Institute and Department of Physics,
University of Chicago, Chicago IL 60637, USA
\newline
$^{ 10}$Fakult\"at f\"ur Physik, Albert Ludwigs Universit\"at,
D-79104 Freiburg, Germany
\newline
$^{ 11}$Physikalisches Institut, Universit\"at
Heidelberg, D-69120 Heidelberg, Germany
\newline
$^{ 12}$Indiana University, Department of Physics,
Swain Hall West 117, Bloomington IN 47405, USA
\newline
$^{ 13}$Queen Mary and Westfield College, University of London,
London E1 4NS, UK
\newline
$^{ 14}$Technische Hochschule Aachen, III Physikalisches Institut,
Sommerfeldstrasse 26-28, D-52056 Aachen, Germany
\newline
$^{ 15}$University College London, London WC1E 6BT, UK
\newline
$^{ 16}$Department of Physics, Schuster Laboratory, The University,
Manchester M13 9PL, UK
\newline
$^{ 17}$Department of Physics, University of Maryland,
College Park, MD 20742, USA
\newline
$^{ 18}$Laboratoire de Physique Nucl\'eaire, Universit\'e de Montr\'eal,
Montr\'eal, Quebec H3C 3J7, Canada
\newline
$^{ 19}$University of Oregon, Department of Physics, Eugene
OR 97403, USA
\newline
$^{ 20}$CLRC Rutherford Appleton Laboratory, Chilton,
Didcot, Oxfordshire OX11 0QX, UK
\newline
$^{ 22}$Department of Physics, Technion-Israel Institute of
Technology, Haifa 32000, Israel
\newline
$^{ 23}$Department of Physics and Astronomy, Tel Aviv University,
Tel Aviv 69978, Israel
\newline
$^{ 24}$International Centre for Elementary Particle Physics and
Department of Physics, University of Tokyo, Tokyo 113, and
Kobe University, Kobe 657, Japan
\newline
$^{ 25}$Institute of Physical and Environmental Sciences,
Brunel University, Uxbridge, Middlesex UB8 3PH, UK
\newline
$^{ 26}$Particle Physics Department, Weizmann Institute of Science,
Rehovot 76100, Israel
\newline
$^{ 27}$Universit\"at Hamburg/DESY, II Institut f\"ur Experimental
Physik, Notkestrasse 85, D-22607 Hamburg, Germany
\newline
$^{ 28}$University of Victoria, Department of Physics, P O Box 3055,
Victoria BC V8W 3P6, Canada
\newline
$^{ 29}$University of British Columbia, Department of Physics,
Vancouver BC V6T 1Z1, Canada
\newline
$^{ 30}$University of Alberta,  Department of Physics,
Edmonton AB T6G 2J1, Canada
\newline
$^{ 31}$Duke University, Dept of Physics,
Durham, NC 27708-0305, USA
\newline
$^{ 32}$Research Institute for Particle and Nuclear Physics,
H-1525 Budapest, P O  Box 49, Hungary
\newline
$^{ 33}$Institute of Nuclear Research,
H-4001 Debrecen, P O  Box 51, Hungary
\newline
$^{ 34}$Ludwigs-Maximilians-Universit\"at M\"unchen,
Sektion Physik, Am Coulombwall 1, D-85748 Garching, Germany
\newline
\bigskip\newline
$^{  a}$ and at TRIUMF, Vancouver, Canada V6T 2A3
\newline
$^{  b}$ and Royal Society University Research Fellow
\newline
$^{  c}$ and Institute of Nuclear Research, Debrecen, Hungary
\newline
$^{  d}$ and Department of Experimental Physics, Lajos Kossuth
University, Debrecen, Hungary
\newline
$^{  e}$ on leave of absence from the University of Freiburg
\newline

\newpage
\section{Introduction}
\label{sect-intro} 

Measurements of event shape variables in hadronic \epem\ annihilations
were amongst the earliest QCD
studies performed at LEP ~\cite{O-shapes,ADL-shapes}.
Such observables may be calculated using perturbative QCD, and are hence
useful for the measurement of the strong coupling
strength, \as.
They are also useful inputs for testing and tuning Monte Carlo models 
of hadronic processes.
An additional test of QCD is provided by the angular dependence of 
these event shape distributions, which arises because of gluon emission.  
However, these effects are small, and therefore require the high 
statistics samples of hadronic \Zzero\ decays which are now available
from the later years of LEP~I running.

At the Born level, with no gluon emission, the spin-1 \Zzero\
created from unpolarised \epem\ beams will decay to spin-\half\ quarks 
with an angular distribution of the form $(1+\cos^2\theta)$ 
where $\theta$ is the angle between the quark and the e$^-$ beam,
together with a parity
violating term proportional to $\cos\theta$.
This simple picture is modified by QCD effects such as the emission of gluons 
and subsequent hadronization.    
The primary quark direction is not directly observable; instead 
the principal event axis can conveniently be specified by the 
thrust axis~\cite{bib-thrust}.  The thrust variable $T$ is defined by
\begin{equation}
   T = \max_{\hat{\mathbf{n}}} \left( \frac{\sum_i|\mathbf{p}_i .  
\hat{\mathbf{n}}|}
       {\sum_i |\mathbf{p}_i|} \right)
\label{eq-thrust}
\end{equation}
where the sum runs over the particles in the event.  
The axis $\hat{\mathbf{n}}$
which maximizes the expression in parentheses is called the thrust axis.
For a three-particle final state, the thrust lies in the range
$[\frac{2}{3},1]$, while for a high multiplicity 
isotropic distribution of particles the range of $T$ 
extends down to $\frac{1}{2}$.
The general form for the distribution of the polar angle of the
thrust axis, \thetaT,  is
\begin{equation} 
   \frac {\dd\sigma} {\dd\costhT} =
   \threeeighths (1+\cos^2\thetaT) \sigmaT + 
   \threequarters \sin^2\thetaT \sigmaL   
\label{eq-diffxs}
\end{equation}
where the parity violating terms are absent, since  
equation~(\ref{eq-thrust}) shows that the 
sense of the thrust axis is arbitrary.
The terms $\sigmaL$ and $\sigmaT$ are referred to as the longitudinal 
and transverse cross-sections respectively; the total cross-section
$\sigmatot=\sigmaL+\sigmaT$.  
The terminology reflects the fact that a longitudinally polarized 
\Zzero, i.e.\ having spin component zero along the \epem\ collision axis,
would yield a $\sin^2\theta$ dependence in its decay to fermions, 
though in the present case this component of the angular distribution 
is being generated entirely by final state QCD radiation.
Note that the longitudinal and transverse cross-sections considered 
here are not the same as the homonymous quantities 
which can be extracted from measurement of fragmentation 
functions in single hadron production~\cite{O-fragl,bib-nw}.
In the present study, we determine \sigmaL\ from an analysis of the 
angular distribution of the thrust axis.  We also study the dependence 
of \sigmaL\ on the value of thrust, $T$.

In ref.~\cite{bib-lampe}, the longitudinal cross-section has been computed
analytically to \oa\ as a function of thrust:
\begin{equation}
\frac{\dd\sigmaL}{\dd T} = \sigma_0 \frac{\as}{2\pi} \CF \frac{2}{T^2}
(8T-3T^2-4) + \oaa  
\label{eq-dsigmaL}
\end{equation}
where $\sigma_0$ is the Born cross-section,
i.e.\ the cross-section in the absence of QCD radiation, and the colour factor
$\CF=\frac{4}{3}$.  In contrast to the overall thrust distribution at \oa,
which diverges as $T \rightarrow 1$~\cite{bib-derujula}, 
the longitudinal cross-section remains finite as $T\rightarrow 1$.
The following prediction for the
longitudinal cross-section has been obtained~\cite{bib-lampe}:
\begin{equation} 
\frac{\sigmaL}{\sigma_0}
=-2(8\ln\twothirds+3)\frac{\as}{2\pi}\CF\left(1+\ell\frac{\as}{2\pi}\right)
+\oaaa
\label{eq-sigmaL}
\end{equation}
where the \oa\ term is obtained by 
integration of equation~(\ref{eq-dsigmaL}), 
and the relative size of the next-to-leading term is 
governed by the value of $\ell=0.7\pm0.2$, obtained from
numerical integration
of the \oaa\ QCD matrix elements.
Note that the \oaa\ contribution is very small, 
of order 1\% of the \oa\ term, leading one to hope that higher order
corrections to \sigmaL\ might also be small, and therefore that a comparison
with data is worthwhile.
In the present study, we actually determine the 
ratio of the longitudinal to the total
cross-section, for which the QCD prediction is:
\begin{equation} 
\frac{\sigmaL}{\sigmatot}=
-2(8\ln\twothirds+3)\frac{\as}{2\pi}\CF\left(1+(\ell-2)\frac{\as}{2\pi}\right)
+\oaaa
\label{eq-RL}
\end{equation}
where we have used the \oa\ expression for the total cross-section,
$\sigmatot=\sigma_0(1+\as/\pi)$.  

The paper is organised as follows:  in Section~\ref{sect-proc} we outline the
experimental procedures adopted, followed by presentation and interpretation of
the results in Section~\ref{sect-results} and finally a brief summary.  

\section{Experimental Procedure}
\label{sect-proc} 
\setcounter{footnote}{2}
The \Opal\ detector has been described in detail 
elsewhere~\cite{bib-opaldet,bib-opalsi}.
For the present analysis, the essential components are the central tracking 
detectors and the electromagnetic calorimeter.
The tracking system consists of a silicon microvertex
detector and three drift chamber systems, all of which
lie within an axial magnetic field of 0.435~T.
The acceptance of the tracking system, with the quality cuts adopted below, 
is roughly $|\costh|<0.93$.\footnote{ 
The \Opal\ coordinate system is defined so that
$z$~is the coordinate parallel to the e$^-$ beam,
$r$~is the coordinate normal to this axis,
$\theta$~is the polar angle with respect to $z$ and
$\phi$~is the azimuthal angle about the $z$-axis.}
The electromagnetic calorimeter is constructed from lead glass blocks,
with a barrel covering $|\costh|<0.81$ and endcaps extending the
acceptance to $|\costh|<0.98$.

Hadronic \Zzero\ decays were selected using standard cuts described
in ref.~\cite{bib-opalline}.  
Tracks to be used in the analysis were selected according to the 
following criteria:  transverse momentum greater than 0.1~GeV/$c$, at least 40
reconstructed points in the main drift chamber, extrapolation to the 
nominal collision point within 2~cm in $r$-$\phi$ and 25~cm in $z$ and
measured momentum less than 65~GeV/$c$.
Energy clusters in the electromagnetic calorimeter were required to 
have at least 0.25~GeV observed energy, and in the endcap region to
contain at least two lead glass blocks.  Background from two-photon and
$\tau^+\tau^-$ events was reduced to a negligible level by demanding at least
seven charged tracks, and the number of poorly contained events was reduced
by demanding $|\costhT|<0.95$.  
With these cuts, approximately 2.1 million events were selected from the 
data recorded in 1993--5, with energies within $\pm 0.5$~GeV of the \Zzero\
peak.  

Simulated events were used to correct the data for the effects of 
detector resolution and acceptance.  The parton shower Monte Carlo
\Jetset~7.4~\cite{bib-jetset}, 
with parameters tuned to \Opal\ data~\cite{O-qg}, was  
used for this purpose.  The events were processed through a 
simulation of the \Opal\ detector~\cite{O-GOPAL}, and reconstructed in the same
way as data.  Approximately 5.6 million simulated events were used in the
analysis.

The thrust value and the direction of the thrust axis were computed from 
the parameters of the observed tracks and clusters.  To account for the 
possible double counting of energy, a standard algorithm~\cite{bib-MT} 
was employed.  In essence, the procedure involved 
removing from each energy cluster in the calorimeter
the expected energy deposition from any associated charged particles.
The experimental resolution on \costhT\ introduced by the detection procedure
is typically around 0.015.

The value of \costhT\ was then histogrammed for all events,
and also in several separate bins of thrust.    
The effects of detector acceptance and resolution were corrected using
a simple bin-by-bin technique.  Each bin in the 
\costhT\ distribution in data was 
multiplied by a correction factor, evaluated from
the ratio between the corresponding distributions in simulated events
at the {\em hadron level} and the {\em detector level}.
The hadron level distribution is computed using the particles 
remaining after those particles having mean 
lifetimes shorter than $3 \times 10^{-10}$~s have decayed.
The detector level distribution is calculated using 
the simulated track and cluster
parameters.                 
Alternatively, the data may be corrected
to the {\em parton level} by using the quarks and gluons 
resulting from the parton shower
instead of the hadrons in the simulated events.
In  general, the parton level results are more appropriate for comparing with
perturbative QCD calculations, while the hadron level results involve less 
model dependence, and can be compared directly with Monte Carlo models 
including hadronization. 

\begin{figure}[ht]
\begin{center}
\resizebox{\textwidth}{!}
{\includegraphics{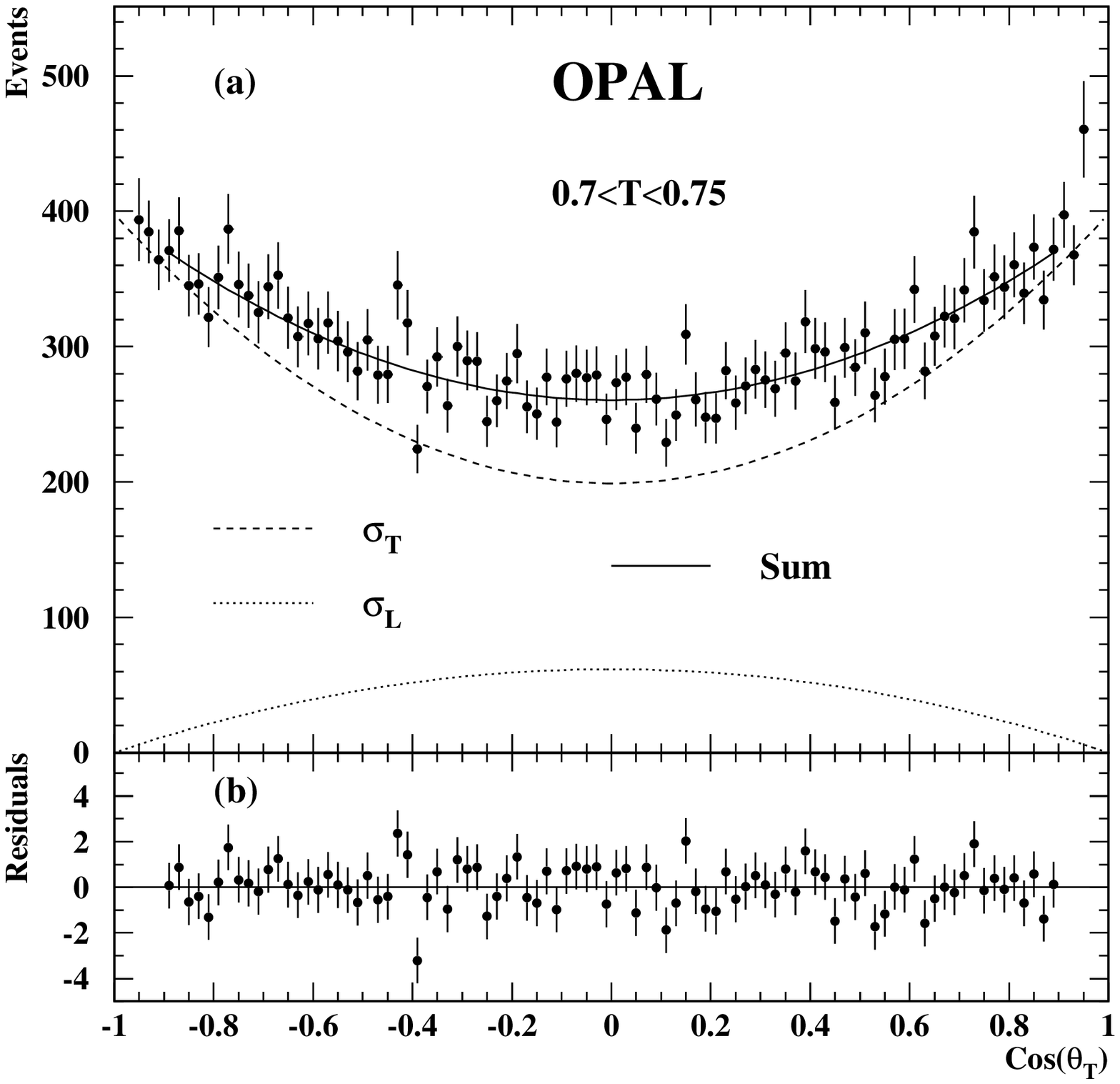}}
\end{center}
\caption[\sl A typical fit to the \costhT\ distribution]
{\sl (a) Fit of equation~(\ref{eqn-fit}) to the corrected data
corresponding to the thrust bin $0.70<T<0.75$; it has $\chi^2$/d.o.f.=79/90. 
The fitted region is $-0.92<\costhT<0.92$. 
The contributions from the longitudinal and transverse
cross-sections are shown separately. (b) The residuals from the fit.}
\label{fig-one}      
\end{figure}

The corrected \costhT\ distributions were then fitted to the form:
\begin{equation}
       A \left( \threeeighths(1-r) (1+\cos^2\thetaT) 
    + \threequarters r \sin^2\thetaT \right)
\label{eqn-fit}
\end{equation}
using a least $\chi^2$ method to determine $r\equiv\sigmaL/\sigmatot$.  
A typical fit to a particular
bin of thrust is shown in fig.~\ref{fig-one}, which yielded 
$r=0.133\pm0.013$.
The values of $\chi^2$ for the fits are good; 
e.g.\ 79 for 90 degrees of freedom in the example shown.

\begin{figure}[ht]
\begin{center}
\resizebox{\textwidth}{!}
{\includegraphics{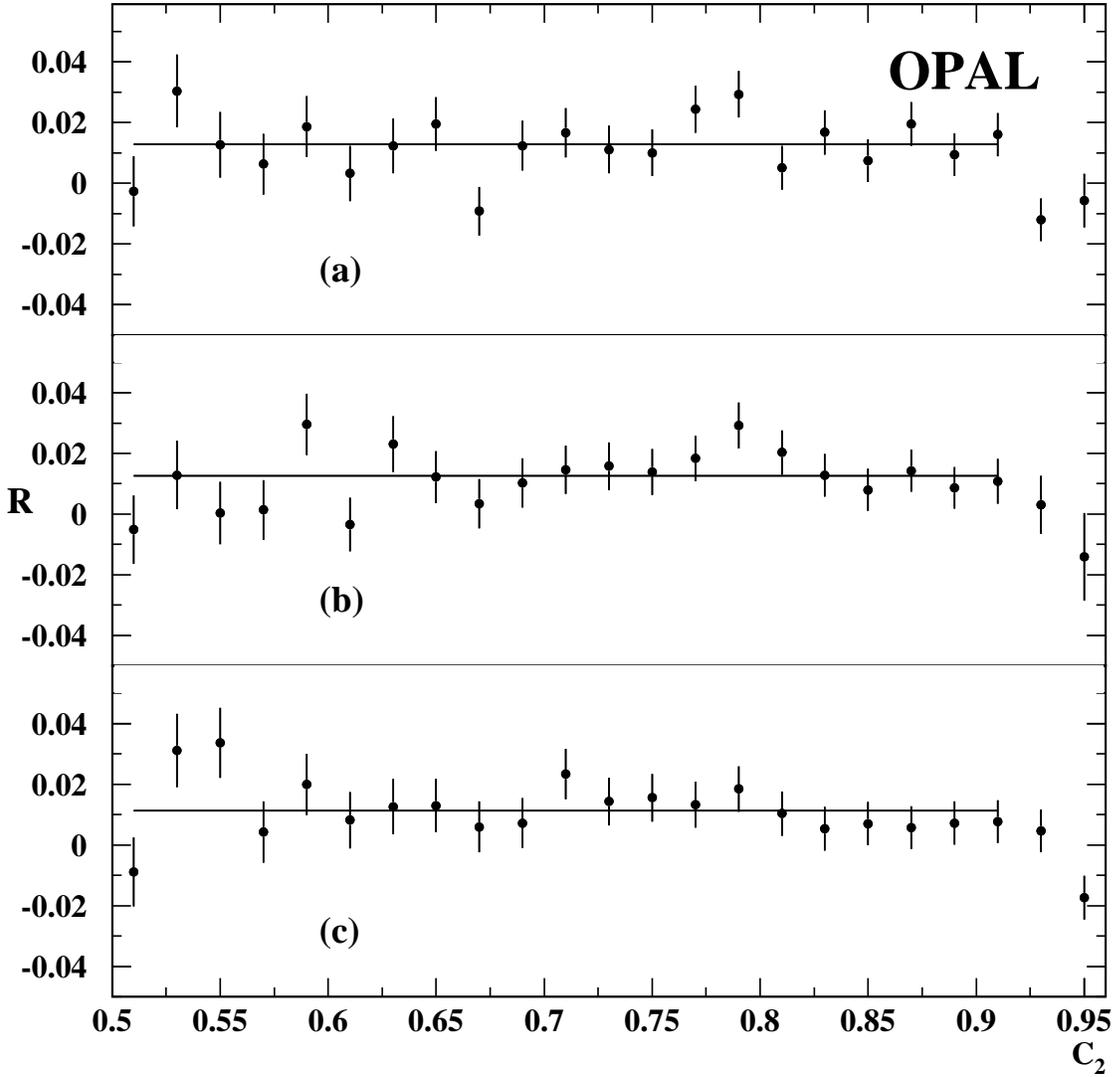}}
\end{center}
\caption[ The]
{\sl 
The ratio $R\approx\sigmaL/\sigmaT$ for all thrust,
computed from pairs of bins in $|\costhT|$,
$C_2$ being the upper value of $|\costhT|$.
The horizontal lines show the average values over the range $0.5<C_2<0.92$.
Three cases are shown: (a) the standard analysis; 
(b) tracks alone; 
(c) calorimetry alone.}
\label{fig-two}
\end{figure}

The choice of the range of \costhT\ in which to fit is important.
The largest possible range should be used in order to minimize the
statistical errors.  However, problems could be anticipated around the region
$0.7<|\costhT|<0.82$, corresponding to the transition between the barrel and
endcap detection systems of \Opal, 
and for $|\costhT|>0.9$, where there is a gap
in acceptance close to the beam directions.  
The correction factors increase rapidly for $|\cos\thetaT|>0.92$, 
and accordingly, we restrict the fits for the determination of \sigmaL\
to the region $|\costhT|<0.92$. 
To check that no other regions of \costhT\ are 
significantly biasing the results,
the data were considered in pairs of narrow bins of width $\pm0.01$ 
centred at $|\costhT|=C_1$
and $|\costhT|=C_2\equiv C_1+0.5$. The corrected numbers of events in 
the two bins, $A_1$ and $A_2$ respectively, can be used to determine 
the ratio 
\begin{equation}
R=\frac{1}{2}\left(\frac{A_1(1+C_2^2)-A_2(1+C_1^2)}{A_2(1-C_1^2)-A_1(1-C_2^2)}
\right) \;\;\;,
\end{equation} 
where $R\approx\sigmaL/\sigmaT=r/(1-r)$,
assuming that the value of the function 
at the centre of the bin is approximately equal to the mean across the bin.     In this way, 25 statistically independent measurements of $R$ were made.
Any anomalous values of $R$ would indicate that the corresponding  
regions of $|\costhT|$ were introducing a bias.  
In Fig.~\ref{fig-two} we show plots of $R$ against $C_2$ for various
combinations of tracking and calorimetry.  
The measurements are seen to be compatible with a constant value for 
$C_2<0.92$.

Systematic errors on \sigmaL\ are assessed by making the following changes
to the analysis:   
\begin{itemize}
\item
Repeat the analysis using 
tracks only and calorimeter clusters only.  
\item
The cuts for selecting charged tracks are modified:
the transverse momentum cut was increased to 0.2~GeV/$c$, 
the cut on reconstructed points was modified to 20
and the cuts on the extrapolation to the 
nominal collision point were tightened to 1~cm in $r$-$\phi$ and 10~cm in $z$,
and the analysis repeated.
\item
Energy clusters in the electromagnetic calorimeter were all required to 
contain at least two lead glass blocks.
\end{itemize}
The changes in the overall longitudinal cross-section under each of these
checks are given in Table~\ref{table-syst}.
The largest change seen in  \sigmaL\ under any of these
variant analyses was taken as the systematic error.
The largest systematic error arose from the use of calorimeter clusters
only, and was found to be associated particularly with the region of 
high thrust, $T>0.95$.
The differences between the longitudinal cross-sections at the parton and 
hadron levels as predicted by the \Jetset\ and \Herwig~\cite{bib-herwig}
models are equal within errors.  Therefore, no systematic error associated
with hadronization was assigned.

\renewcommand{\arraystretch}{1.2}
\begin{table}[ht]
\begin{center}
\begin{tabular}{lr}
\hline
Change in analysis & $\Delta(\sigmaL/\sigmatot)$ \\
\hline
Tracks only & $+0.0002$ \\
Calorimetry only & $-0.0013$ \\
Track $r-\phi$ cut & $+0.0005$ \\
Track $z$ cut & $+0.0006$ \\
Track points cut & $-0.0006$ \\
Track $p_{\mathrm{T}}$ cut & $-0.0003$ \\
Cluster blocks cut & $0.0005$ \\
\hline
\end{tabular}
\end{center}
\caption[\sl ]
{\sl  Changes in the measured $\sigmaL/\sigmatot$ (the same at both 
parton and hadron 
levels) under various systematic changes in the analysis.}
\label{table-syst}      
\end{table}

\section{Results}
\label{sect-results} 

The values found for \sigmaL\ in the data, integrated over all thrust,
corrected to the parton and hadron level respectively are:
\begin{eqnarray*}
\frac{\sigmaL}{\sigmatot}=0.0127\pm0.0016(\mathrm{stat.})
\pm0.0013(\mathrm{syst.}) & \;\;\;\;\; & \mathrm{(Parton\; level)} \\
\frac{\sigmaL}{\sigmatot}=0.0121\pm0.0016(\mathrm{stat.})
\pm0.0013(\mathrm{syst.}) & \;\;\;\;\; & \mathrm{(Hadron\; level)} \\
\end{eqnarray*}
The difference between these values reflects the influence of hadronization.
We note that the values corrected to the parton and hadron levels are 
very close, showing that the effect of hadronization on \sigmaL/\sigmatot\ is, 
according to the \Jetset\ model, rather small ($\sim -5\%$).
An essentially identical difference between parton and hadron levels
is predicted by the \Herwig\  model.
A current world average value for the strong coupling is~\cite{bib-pdg}
$\amz=0.119\pm0.002$.  Using equation~(\ref{eq-RL}), this 
yields a prediction at \oaa\ of $\sigmaL/\sigmatot=0.0120\pm0.0002$, 
where the error arises predominantly from the uncertainty in \amz.  
This prediction is most appropriately compared with the parton level
measurement, which is in excellent agreement with the expectation
from \oaa\ QCD.  

\begin{figure}[ht] 
\begin{center}
\resizebox{\textwidth}{!}
{\includegraphics{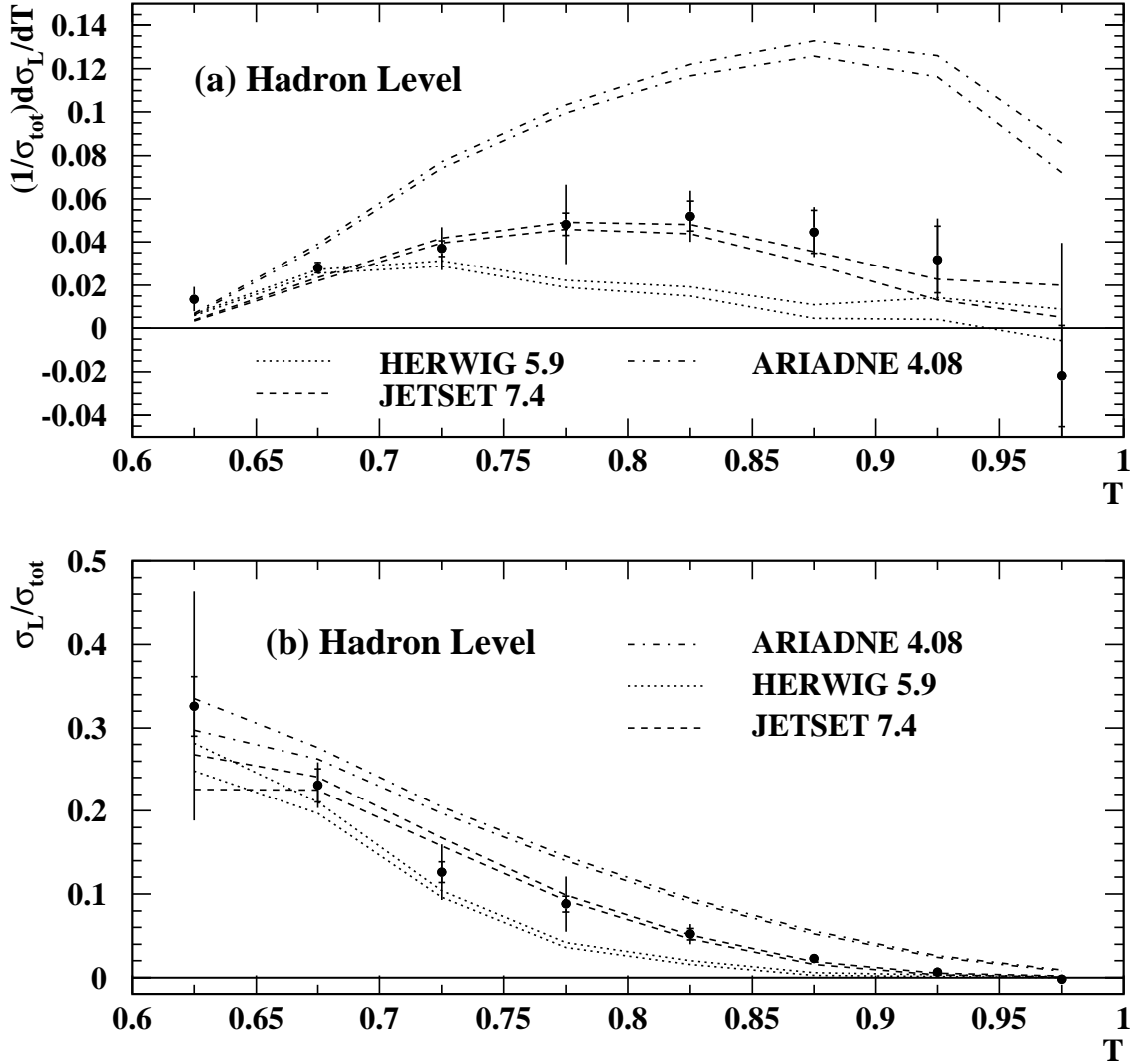}}
\end{center}
\caption[]
{(a) $(1/\sigmatot)(\dd\sigmaL / \dd T)$ for data corrected to the hadron
level.  The cross-marks on the error bars show the statistical errors.
The lines indicate the predictions of the \Jetset, \Herwig\
and \Ariadne\
models, with the width between them 
indicating the statistical uncertainties arising from samples of
$10^7$ events.
(b) Ratio of the longitudinal to total cross-sections, 
$\sigmaL/\sigmatot$, 
in each bin of $T$ for data corrected to the hadron
level.  The cross-marks on the error bars show the statistical errors.
The lines again indicate the predictions of the \Jetset, \Herwig\
and \Ariadne\ models.
}
\label{fig-three}
\end{figure}

The measured values for the differential cross-section, 
$(1/\sigmatot)\dd\sigmaL/\dd T$, corrected to the hadron level, are presented
in Table~\ref{table-one} and Fig.~\ref{fig-three}(a).
Systematic errors have been assessed using the same method as 
for the overall longitudinal cross-section.   
In contrast to the overall differential thrust distribution, which is 
strongly peaked towards $T=1$ corresponding to two-jet events, 
the longitudinal cross-section displays a broad distribution.  
The statistical errors on the longitudinal cross-section become 
larger as $T \rightarrow 1$, reflecting the fact that the longitudinal
cross-section becomes a smaller proportion of the total.  
The measured value for $0.95<T<1$ is negative, though consistent with
zero within statistical errors.   
In Fig.~\ref{fig-three}(b) we show the longitudinal to total cross-section
ratio, $r$, as a function of thrust.  The increasing importance of the
longitudinal component of the cross-section as thrust decreases is apparent.
The ratio is seen to approach the value \onethird, corresponding to isotropic  
orientation of the thrust axis (c.f.~equation~(\ref{eq-diffxs})), for the
lowest values of $T$. 

\begin{table}[ht]
\begin{center}
\begin{tabular}{c|r@{$\pm$}c@{$\pm$}c|r@{$\pm$}c@{$\pm$}c}
\hline
$T$ & \multicolumn{3}{c}
{\rule[-4mm]{0pt}{11mm} $\frac{1}{\sigmatot}\frac{\dd\sigmaL}{\dd T}$} 
&  \multicolumn{3}{|c} {$r=\frac{\sigmaL}{\sigmatot}$}  \\
\hline
0.60--0.65 & 0.013   & $0.002$ & $0.005$ & 0.33 & 0.04 & 0.13 \\
0.65--0.70 & 0.028   & $0.002$ & $0.001$ & 0.231 & 0.020 & 0.019 \\
0.70--0.75 & 0.037   & $0.004$ & $0.009$ & 0.126 & 0.013 & 0.031 \\
0.75--0.80 & 0.048   & $0.005$ & $0.018$ & 0.088 & 0.009 & 0.032 \\
0.80--0.85 & 0.052   & $0.009$ & $0.010$ & 0.052 & 0.007 & 0.010 \\
0.85--0.90 & 0.045   & $0.010$ & $0.006$ & 0.023 & 0.005 & 0.003 \\
0.90--0.95 & 0.032   & $0.016$ & $0.011$ & 0.007 & 0.003 & 0.002 \\
0.95--1.0  &$-$0.022 & $0.023$ & $0.057$ & $-$0.002 & 0.002 & 0.005 \\
\hline
\end{tabular}
\end{center}
\caption[\sl ]
{\sl  Measured values for the differential cross-section, 
$(1/\sigmatot)\dd\sigmaL/\dd T$, corrected to the hadron level, and 
for the corresponding ratio of longitudinal to total cross-sections.
The first error is statistical and the second systematic.}
\label{table-one}      
\end{table}

We also show in Figs.~\ref{fig-three}(a,b) the predictions of the 
\Jetset~\cite{bib-jetset} (version 7.4), \Herwig~\cite{bib-herwig} 
(version 5.9) and \Ariadne~\cite{bib-ariadne} (version 4.08)
parton shower Monte Carlo models.  
The parameter sets for \Jetset\ and \Herwig\ are obtained from 
fitting to \Opal\ data as described in ref.~\cite{O-qg}, 
except that for \Herwig\
the cluster mass cutoff \verb|CLMAX| was increased to 3.75~GeV
to improve the modelling of the average charged multiplicity, and the
parameters
for \Ariadne\ are taken from the \Opal\ tuning of ref.~\cite{O-shapes}.
The model predictions 
have limited statistical precision, so they are represented by bands
indicating their uncertainties.   Both parts of Fig.~\ref{fig-three}
show similar effects, as expected because a description of the overall
thrust distribution was one of the constraints used in tuning the models.
The \Jetset\ prediction describes the 
data well, and indicates that the longitudinal cross-section tends to
fall towards $T=1$.  The \Herwig\ prediction is
in less good agreement with the data, 
particularly in the intermediate region around $T\sim0.8$.
The \Ariadne\ model, with the parameters used here, gives a sizeable
overestimate of the longitudinal cross-section at almost all values of $T$.
Other parameter sets for \Ariadne, e.g. the defaults or those
given in ref.~\cite{A-physrep}, show essentially the same behaviour.
The overall values of $\sigmaL/\sigmatot$ at the hadron level 
for the three models are 0.0112 (\Jetset), 0.0060 (\Herwig) 
and 0.0335 (\Ariadne), showing again that only \Jetset\ is compatible with 
the \Opal\ data.

\begin{figure}[ht] 
\begin{center}
\resizebox{\textwidth}{!}
{\includegraphics{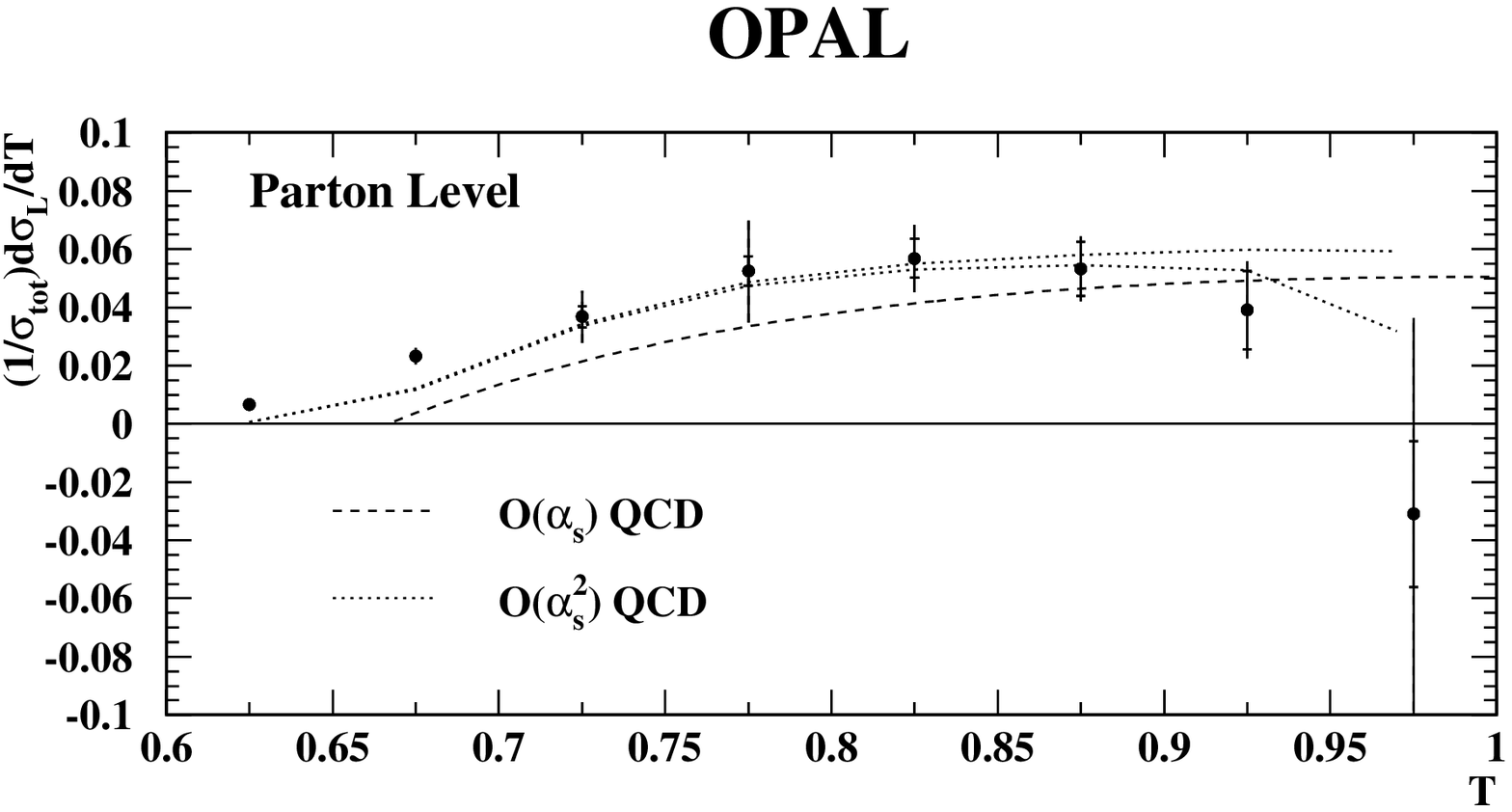}}
\end{center}
\caption[]
{$(1/\sigmatot)(\dd\sigmaL / \dd T)$ for data corrected to the parton
level.  The cross-marks on the error bars show the statistical errors.
The dashed and dotted lines show the \oa\ and \oaa\ QCD predictions
respectively, taking \amz=0.119.
In the latter case, the statistical error on the prediction 
is indicated by the width of the band.  The errors on the \oaa\ predictions
diverge for $T\rightarrow 1$, and so the value plotted at $T=0.97$ represents
an average over the range $0.95<T<0.99$, while the nearby data point
covers the bin $0.95<T<1$.}
\label{fig-four}
\end{figure}

The measured values for the differential cross-section, 
$(1/\sigmatot)\dd\sigmaL/\dd T$, corrected to the parton level, are presented
in Fig.~\ref{fig-four}.   The differences with respect to the hadron level 
distribution (Fig.~\ref{fig-three}(a)) are small.   
The prediction of \oa\ QCD (equation~(\ref{eq-dsigmaL})),
taking \amz=0.119, is shown by the dashed line.
The distribution as a function of thrust
is poorly described, with the \oa\ calculation significantly 
underestimating the data at low thrust. 
This is qualitatively the
behaviour predicted in ref.~\cite{bib-lampe}, and contrasts with the
case of the longitudinal cross-section integrated over thrust,   
where the \oa\ and \oaa\ calculations are very close, and in good agreement
with data, as noted above.
The \oaa\ QCD prediction for the differential cross-section
may be obtained numerically using the program
\verb|EVENT2|~\cite{bib-event2}, and is indicated by the dotted band in 
Fig.~\ref{fig-four}.  This \oaa\ calculation gives a much improved
prediction of  $(1/\sigmatot)\dd\sigmaL/\dd T$, though the data 
still lie above the prediction for the lowest thrust values, $T<0.7$.
At high values of $T$, especially above $\sim 0.97$, 
the errors on the \verb|EVENT2|
predictions (and on the data) become too large 
for any clear conclusion to be drawn, and in consequence 
we have not been able to use \verb|EVENT2| to validate the \oaa\ calculation
of ref.~\cite{bib-lampe}.
 
\section{Summary}
\label{sect-summ} 

In this paper, we have presented a determination of the longitudinal 
cross-section in hadronic electron-positron annihilations on the \Zzero\
peak.  The values obtained:
\begin{eqnarray*}
\frac{\sigmaL}{\sigmatot}=0.0127\pm0.0016(\mathrm{stat.})
\pm0.0013(\mathrm{syst.}) & \;\;\;\;\; & \mathrm{(Parton\; level)} \\
\frac{\sigmaL}{\sigmatot}=0.0121\pm0.0016(\mathrm{stat.})
\pm0.0013(\mathrm{syst.}) & \;\;\;\;\; & \mathrm{(Hadron\; level)} \\
\end{eqnarray*}
are in good agreement with the prediction of \oaa\ QCD, and also with 
the only previously published measurement of this quantity~\cite{A-physrep}.
The dependence of the longitudinal cross-section on thrust has been
investigated.  Good agreement with the \Jetset\ parton shower Monte Carlo model
is observed.  The \Herwig\ and \Ariadne\ models, with the parameter set tuned
to \Opal\ data, show less good agreement. 
Comparing with fixed order QCD predictions, we find that the 
\oa\ QCD prediction differs significantly from the data,
while an \oaa\ calculation shows much better agreement.  Thus, in contrast to
the integrated longitudinal cross-section, the differential cross-section 
exhibits significant higher order effects.

\section{Acknowledgements}
\label{sect-ack} 
\par
We particularly wish to thank the SL Division for the efficient operation
of the LEP accelerator at all energies
 and for their continuing close cooperation with
our experimental group.  We thank our colleagues from CEA, DAPNIA/SPP,
CE-Saclay for their efforts over the years on the time-of-flight and trigger
systems which we continue to use.  In addition to the support staff at our own
institutions we are pleased to acknowledge the  \\
Department of Energy, USA, \\
National Science Foundation, USA, \\
Particle Physics and Astronomy Research Council, UK, \\
Natural Sciences and Engineering Research Council, Canada, \\
Israel Science Foundation, administered by the Israel
Academy of Science and Humanities, \\
Minerva Gesellschaft, \\
Benoziyo Center for High Energy Physics,\\
Japanese Ministry of Education, Science and Culture (the
Monbusho) and a grant under the Monbusho International
Science Research Program,\\
German Israeli Bi-national Science Foundation (GIF), \\
Bundesministerium f\"ur Bildung, Wissenschaft,
Forschung und Technologie, Germany, \\
National Research Council of Canada, \\
Research Corporation, USA,\\
Hungarian Foundation for Scientific Research, OTKA T-016660, 
T023793 and OTKA F-023259.\\


\end{document}